\documentclass[conference]{IEEEtran}
\IEEEoverridecommandlockouts
\usepackage{cite}
\usepackage{url}
\usepackage{amsmath,amssymb,amsfonts}
\usepackage{algorithmic}
\usepackage{graphicx}
\usepackage{textcomp}
\usepackage{xcolor}
\usepackage{amssymb}
\usepackage{multirow}
\usepackage{bbding} 
\usepackage{hyperref}

\usepackage{algorithm}
\usepackage{colortbl}
\usepackage{diagbox}
\usepackage{booktabs} 
\usepackage{tabularx}
\usepackage{float}
\def\BibTeX{{\rm B\kern-.05em{\sc i\kern-.025em b}\kern-.08em
    T\kern-.1667em\lower.7ex\hbox{E}\kern-.125emX}}
\begin{document}

\title{MoMuSE: Momentum Multi-modal Target Speaker
Extraction for Real-time Scenarios with Impaired Visual Cues 
\thanks{This work is supported by China NSFC projects under Grants No. 62401377 and 62271432.}
}

\author{Junjie Li$^1$, Ke Zhang$^{2}$, Shuai Wang$^{2,3,*}$,  Kong Aik Lee$^{1,*}$, Man-Wai MAK$^1$, Haizhou Li$^{2,3}$\thanks{$^*$: Corresponding author.} \\
\IEEEauthorblockN{
$^1$ Department of Electrical and Electronic Engineering, The Hong Kong Polytechnic University, Hong Kong SAR\\ 
$^2$ Shenzhen Research Institute of Big Data, Shenzhen, China \\
$^3$ School of Data Science, The Chinese University of Hong Kong, Shenzhen (CUHK-Shenzhen), China 
}
}
\maketitle

\begin{abstract}
Audio-visual Target Speaker Extraction (AV-TSE) aims to isolate the speech of a specific target speaker from an audio mixture using time-synchronized visual cues. In real-world scenarios, visual cues are not always available due to various impairments, which undermines the stability of AV-TSE. Despite this challenge, humans can maintain attentional momentum over time, even when the target speaker is not visible.
In this paper, we introduce  the \textit{Momentum Multi-modal target Speaker Extraction} (MoMuSE), which retains a speaker identity momentum in memory, enabling the model to continuously track the target speaker. Designed for real-time inference, MoMuSE extracts the current speech window with guidance from both visual cues and dynamically updated speaker momentum. Experimental results demonstrate that MoMuSE exhibits significant improvement, particularly in scenarios with severe impairment of visual cues.
\end{abstract}


\vspace{-2mm}
\section{Introduction}

\label{sec:intro}
Audio-visual Target Speaker Extraction (AV-TSE)~\cite{gao2021visualvoice,li2024audio,tao2024audio} aims to extract the speech of a target speaker using visual cues, such as lip or face image sequences, as references. AV-TSE, known for their resilience against acoustic noise, have demonstrated superior performance compared to audio-only approaches \cite{wang2020robust,10446297} in complex acoustic environments.

Despite the significant benefits that visual information provides to AV-TSE tasks, challenges arise when visual cues are temporarily unavailable in real scenarios, such as  absence of the target person's video. Additionally, issues like lip occlusion or a low-quality camera~\cite{afouras19_interspeech} can result in unclear lip areas. These challenges make AV-TSE systems relying on time-aligned sequences impractical under real-world environments.

Several related studies have tried to address the aforementioned challenges. Sadeghi and Alameda-Pineda \cite{sadeghi2021switching,sadeghi2020robust} propose switching from an audio-visual variational auto-encoder (VAE) to an audio-only VAE when visual quality is poor. Wu et al. \cite{wu2022time} incorporate an attention mechanism, selecting relevant visual features based on mixed audio features. In contrast to discarding low-quality visual features, Pan et al. \cite{pan2023imaginenet} adopt an innovative approach by reconstructing corrupted video through audio-visual correspondence. Furthermore, the VS model \cite{afouras19_interspeech} and the audio-visual SpeakerBeam \cite{sato2021multimodal,ochiai2019multimodal} learn a speaker embedding from pre-enrolled speech and rely on this embedding when visual cues are unreliable. Liu et al. \cite{10596551} propose a triple loss function, where visual cues are leveraged only during the training phase and are not directly fused into the model’s input features.

Despite achieving good performance, above approaches introduce an additional pre-enrollment step, and the acoustic environments of the pre-enrolled speech may not match the deployment environments.  Instead of pre-enrollment, we propose tracking speaker identity using speech derived from previous time steps, offering a more accurate representation of the speaker's current vocal characteristics. While leveraging such ``self-enrollment" speech for second-phase inference has proven effective in tasks like speech enhancement \cite{andreev23_interspeech}, audio-based speech separation \cite{wang2020online,li2018source,li2019listening,pan24_interspeech}, and target speaker extraction \cite{deng21c_interspeech,pan2024neuroheed,pandey2023attentive}, the performance degrades if the self-enrollment speech is of poor quality.

In this paper, we propose  \textit{\textbf{Mo}mentum \textbf{Mu}lti-modal target \textbf{S}peaker \textbf{E}xtraction} (\textbf{MoMuSE}) \footnote{Demo: \url{https://mrjunjieli.github.io/demo_page/MoMuSE/index.html}},  a novel framework that employs a memory bank with a momentum mechanism to dynamically track and update the speaker's identity using previous time steps, This dynamic updating ensures that the memory bank retains the most accurate speaker embedding, enabling MoMuSE to maintain focus on the target speaker during real-time, even when visual cues are corrupted. Experiments demonstrate that our proposed MoMuSE achieves a momentum effect, enabling TSE to continue functioning even when visual cues are completely absent.

\vspace{-1mm}
\section{Visual Impairments}

\begin{figure}[ht]
\vspace{-4mm}
    \centering
    \includegraphics[width=0.4\textwidth]{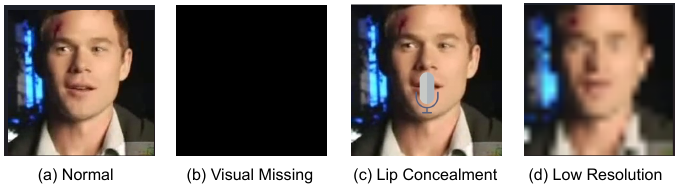}
    \vspace{-4mm}
    \caption{Examples of normal and impaired visual frames. }
    \label{fig:impariment}
     \vspace{-3mm}
\end{figure}
\vspace{-1mm}

\begin{figure*}[htbp]
    \centering
    \includegraphics[width=0.92\textwidth]{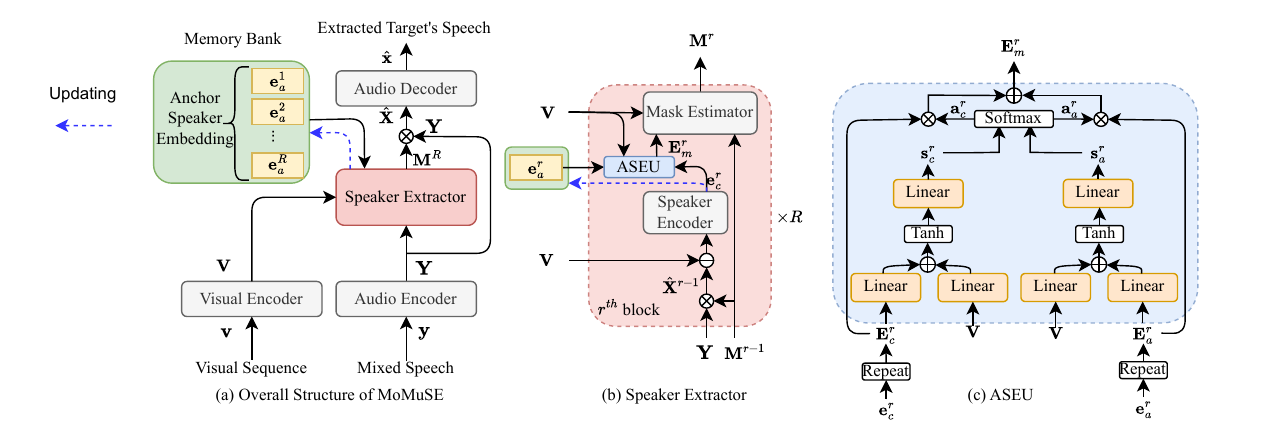}
    \vspace{-5mm}
    \caption{(a) Overall structure of MoMuSE. Modules in gray denote  the original structures from MuSE, while modules in other colors denote the  proposed  new structures.  (b) Speaker Extractor.
    (c) The detailed structure of \textit{\textbf{A}nchor \textbf{S}peaker \textbf{E}mbedding \textbf{U}pdating} (\textbf{ASEU}), which is an attention based module.   $\otimes$, $\ominus$ and $\oplus$ refer to point-wise multiplication, concatenation and point-wise addition.  }
    \label{fig:Model}
    \vspace{-5mm}
\end{figure*}


 The normal visual frame is depicted in Fig. \ref{fig:impariment}(a). However, in real-world scenarios, clear visual frames may not be available throughout entire conversations. As shown in Fig.~\ref{fig:impariment}, prevalent issues include: 1) \textbf{Visual Missing}, where the speaker's face remains undetected; 2) \textbf{Lip Concealment}, where objects such as hands or microphones obscure the mouth region; and 3) \textbf{Low Resolution}, caused by inferior camera quality, the camera being out of focus, or the speaker's distance from the camera.

\vspace{-1mm}
\section{Proposed Method}

To enhance the stability of AV-TSE systems in the visual impairment scenarios and avoid introducing a pre-enrollment process \cite{10631297,wang2024wesep,zhang2024multi}, we  leverage the advantages of dynamic visual frames when visual cues are reliable while incorporating alternative speaker voiceprint cue when visual information is impaired. In this paper, we introduce MoMuSE, a novel framework that employs a dynamically updated speaker momentum as an additional reference. MoMuSE enables robust speaker extraction even in the presence of visual impairments, presenting a more effective way to harness the complementary nature of visual and audio modalities.

\vspace{-1mm}
\subsection{Recap of MuSE}
MuSE \cite{pan2021muse} is an AV-TSE model that utilizes visual cues and a self-enrolled speaker embedding  as references. As depicted in Fig \ref{fig:Model}, the modules in gray represent the original MuSE structure, while the modules in other colors indicate our proposed new components.

MuSE comprises four key modules: an audio encoder, an audio decoder, a visual encoder, and a speaker extractor. The audio encoder converts the time-domain mixed speech $\mathbf{y}$ into a latent representation $\mathbf{Y}$, and the audio decoder reconstructs the extracted target speech $\mathbf{x}$ from its latent representation $\hat{\mathbf{X}}$. The visual encoder processes the visual sequence $\mathbf{v}$ to generate visual embeddings $\mathbf{V}$. The speaker extractor iteratively refines the quality of the extracted speech using $R$ extractor blocks. Each block includes a speaker encoder and a mask estimator, as illustrated in Fig \ref{fig:Model} (b). The speaker encoder fuses the estimated speech representation $\hat{\mathbf{X}}^{r-1}$ and the visual representation $\mathbf{V}$ to generate a speaker embedding $\mathbf{e}^r_c$. The mask estimator then generates a mask $\mathbf{M}^r$ to isolate the target speech. Here, $r=(1, 2, \ldots, R)$ indicates the index of the extractor block, with distinct parameter 
weights for each block.

In MuSE, the self-enrolled speaker embedding $\mathbf{e}^r_c$ is derived from the current window of visual and audio representations, assuming the visual information is consistently reliable. However, this assumption fails under visual impairments.
Our proposed MoMuSE addresses this limitation by incorporating a dynamic speaker momentum mechanism. It leverages a memory bank to store historical information, updating embeddings based on a reliability assessment of the visual cues, thus enhancing stability even when visual inputs are absent.

\vspace{-1mm}
\subsection{MoMuSE Architecture}
\label{momuse}
MoMuSE shares a similar structure with MuSE, yet it incorporates a memory bank that allows it to leverage historical voiceprint information from previously extracted speech, as illustrated in Fig \ref{fig:Model}(a).
The memory bank comprises $R$ anchor speaker embeddings, denoted as ${\mathbf{e}^1_a, \dots, \mathbf{e}^R_a}$, which store the historical voiceprint information of the target speaker. Here, the subscript $a$ denotes ``anchor''. In addition, to effectively combine historical and current speaker information, we propose an \textit{\textbf{A}nchor \textbf{S}peaker \textbf{E}mbedding \textbf{U}pdating} (\textbf{ASEU}) sub-module within each speaker extractor block. This sub-module integrates historical embeddings into the extraction process for the current window, adapting based on the visual features' quality in the current window.
We implement the ASEU sub-module using an additive attention fusion\footnote{\url{https://github.com/sooftware/attentions/blob/master/attentions.py}} mechanism inspired by \cite{sato2021multimodal,DBLP:journals/corr/BahdanauCB14}, as shown in Fig \ref{fig:Model}(c).

To maintain temporal resolution consistency, we apply temporal replication to the speaker embeddings\footnote{For notational clarity, the superscript $r$ denoting the $r$-th block will be neglected in Section \ref{momuse} and Section \ref{momentum}.}
\begin{align}
\mathbf{E}_\psi \in \mathbb{R}^{H \times L} = \text{Repeat}(\mathbf{e}_\psi \in \mathbb{R}^{H \times 1}),
\end{align}
where $H$ and $L$ are the feature dimension and time length, respectively, and  $\psi \in \{a, c\}$ denotes the specific type of embedding being used. Next, the attention mechanism computes two score vectors $\mathbf{s}_c $ and $\mathbf{s}_a$ 
for $\mathbf{E}_c$ and $\mathbf{E}_a$, respectively:
\begin{align}
    \mathbf{s}_\psi \in \mathbb{R}^{1\times L}&= \text{Linear}(\text{Tanh}(\text{Linear}(\mathbf{E}_\psi)+\text{Linear}(\mathbf{V}))),
\end{align}
where $\mathbf{V}\in \mathbb{R}^{H\times L}$ represents the visual feature. These score vectors are then normalized using a softmax function to produce element-wise attention weights: 
\begin{equation}
    \mathbf{a}_{\psi,l} \in \mathbb{R}^{1} =\frac{\exp(\mathbf{s}_{\psi,l})}{\exp(\mathbf{s}_{c,l})+\exp(\mathbf{s}_{a,l})} \quad  (l = 1, 2, \ldots, L),
\end{equation}
where $l$ denotes the time step. Finally, the momentum-based speaker embedding sequence $\mathbf{E}_m$ is computed by fusing $\mathbf{E}_c$ and $\mathbf{E}_a$ according to their attention weights:
\begin{equation}
    \mathbf{E}_m \in \mathbb{R}^{H\times L}= \mathbf{a}_c \otimes \mathbf{E}_c +  \mathbf{a}_a \otimes \mathbf{E}_a ,
\end{equation}
where $\otimes$ denotes point-wise multiplication. In theory, when the current visual feature $\mathbf{V}$ is unreliable, the attention mechanism tends to assign lower weights to current speaker embedding sequence $\mathbf{E}_c$. 
\vspace{-1mm}
\subsection{Momentum Mechanism}
\label{momentum}
In online processing \cite{pan2024neuroheed}, the model begins processing only after the incoming mixed input speech accumulates to a predefined initialization length $L_\text{init}$ (in seconds). Once this initialization is complete, the model processes the mixed speech $\mathbf{y}$ using a window size of $L_\text{win}$ (in seconds), and then advances forward by a step size of $L_\text{shift}$ (in seconds). At the $t$-th window step\footnote{$t$ denotes the window step, while $l$ indicates the time step. Each windows step $t$ contains multiple time steps.}, the model takes the mixed speech $\mathbf{y}(t)$ and the corresponding visual sequence $\mathbf{v}(t)$ as inputs, and outputs the estimated target speech $\hat{\mathbf{x}}(t)$:
\begin{equation}
    \hat{\mathbf{x}}(t) = \text{MoMuSE}(\mathbf{y}(t),\mathbf{v}(t)).
\end{equation}

The momentum mechanism is implemented in online scenarios and contains three components: \textbf{initialization}, \textbf{generation}, and \textbf{updating}.

\textbf{Initialization}: To maintain continuous attention on the target speaker in unreliable visual scenarios, the historical voiceprint information of the target speaker needs to be preserved. We introduce a memory bank to store this information. At the first window step (initialization), the memory bank is empty. The dimension of the mixed speech of the current window is  $\mathbf{y}(1) \in \mathbb{R}^{1 \times L_\text{init}}$. Given that the memory bank is initially empty, the ASEU sub-module is not utilized at this step. Instead, the current speaker embedding sequence $\mathbf{E}_c(1)$ is directly used as input to the mask estimator. Once the corresponding target speech is estimated, the memory bank is initialized with the current speaker embedding $\mathbf{e}_c(1)$:
\begin{equation} \mathbf{e}_a = \mathbf{e}_c(1). \label{eq:eq1}\end{equation}
This initialization allows the memory bank to begin tracking the target speaker’s voiceprint for subsequent processing steps.

\textbf{Generation}: In subsequent window steps ($t > 1$), MoMuSE leverages the ASEU sub-module to effectively balance the historical and current voiceprint information. Using the visual representation $\mathbf{V}(t)$ as the key\footnote{$\mathbf{V}(t)$ is the visual feature  corresponding to the input visual sequence $\mathbf{v}(t)$.}, the model generates a momentum-based speaker embedding sequence $\mathbf{E}_m(t)$.

The attention mechanism within the ASEU computes varying attention weights at each time step, dynamically adjusting the contribution of the current speaker embedding sequence $\mathbf{E}_c(t)$ and the historical anchor embedding sequence $\mathbf{E}_a(t)$. This process allows MoMuSE to adaptively fuse current and past voiceprint information, ensuring robust target speaker extraction even when visual features are impaired or unreliable

\textbf{Updating}: To ensure only high-quality speaker representation is retained in the memory bank, updates are performed when a superior speaker embedding is detected. The average attention weights $\overline{\mathbf{a}}_c(t)$ is employed as the criterion to determine whether a memory update should be triggered: 
\begin{equation}
\mathbf{e}_a =
\begin{cases}
\mathbf{e}_c(t) & \text{if } \overline{\mathbf{a}}_c(t) > \theta  \text{ for } t=2,3,4,…,T\\
\mathbf{e}_a & \text{otherwise},
\end{cases}
\end{equation}
where $\overline{\mathbf{a}}_c(t)  \in \mathbb{R}^{1}$ is the mean of the attention weights $\mathbf{a}_c(t)$ across the time dimension, and $\theta$ is a predefined threshold.

\vspace{-1mm}
\subsection{Loss Function}
Following MuSE \cite{pan2021muse}, we use the scale-invariant signal-to-noise ratio (SI-SNR) loss \cite{luo2019conv} and cross-entropy (CE) loss:
\begin{equation}
    \mathcal{L}_{\text{SI-SNR}}(\mathbf{x},\hat{\mathbf{x}}) = -10 \log_{10}\frac{||\frac{<\hat{\mathbf{x}},\mathbf{x}>\mathbf{x}}{||\mathbf{x}||^2}||^2}{||\hat{\mathbf{x}}-\frac{<\hat{\mathbf{x}},\mathbf{x}>\mathbf{x}}{||\mathbf{x}||^2}||},
    \vspace{-1mm}
\end{equation}
\begin{equation}
    \mathcal{L}_{\text{CE}}(\mathbf{e}^r_c) = -\sum^N_{n=1}d_n \log(\operatorname{softmax}(\mathbf{W}^r \mathbf{e}^r_c)),
    \vspace{-1mm}
\end{equation}
where $d_n$  is class label for speaker $n$ and $\mathbf{x}$ is target speech. $N$ is the number of training speakers. $\mathbf{W}^r$ is a learnable weight matrix 
 in the $r$-th block used to project speaker embedding $\mathbf{e}^r_c$ to a one-hot class label. 

Additionally, we propose a penalty loss on the attention weight $\mathbf{a}^r_a$, mitigating the model's excessive reliance on the anchor speaker embedding $\mathbf{e}^r_a$:
\begin{equation}
    \mathcal{L}_{\text{pe}} = \sum^R_{r=1}\frac{||\mathbf{a}^{r}_a||_1 }{L}.
    \vspace{-1mm}
\end{equation}
Since the current speaker embedding $\mathbf{e}^r_c$ is more relevant to the current speech utterance, we encourage the model to learn more from it, especially when $\mathbf{e}^r_c$ is reliable.

 \vspace{-1mm}
\subsection{Training Procedure} 
In autoregressive training, the past extracted speech is used as input for the current extraction, which can significantly increase computational overhead if every sliding window is optimized. Inspired by the training settings in NeuroHeed \cite{pan2024neuroheed}, we propose a Segment-Level Optimization (\textbf{Seg}) strategy that simulates only the first two window steps during training. This approach effectively reduces computational cost while still capturing the essence of the autoregressive process.

To further enhance model convergence, we introduce two additional strategies: Parameter Initialization (\textbf{PI}), which helps the model start with better initial weights, and Utterance-Level Optimization (\textbf{Utt}), which focuses on optimizing the model at the utterance level to refine overall performance.

     \textbf{PI}: Before training MoMuSE, we initialize its parameters (excluding \textbf{ASEU}) using the checkpoint from the $50$-th training epoch of MuSE.
     
     \textbf{Utt}: MoMuSE processes the whole utterance of visual and audio inputs, ensuring  the model encounters long utterance, which aims to improve the robustness of the speaker encoder: 
    \begin{align}
        \hat{\mathbf{x}}_{\text{}},\mathbf{e}^r_c &= \text{MoMuSE}(\mathbf{y}, \mathbf{v}),  \\
        \mathcal{L}_{\text{utt}}&=\mathcal{L}_{\text{SI-SNR}}(\mathbf{x},\hat{\mathbf{x}}_{\text{}})+\lambda \sum^R_{r=1}\mathcal{L}_{\text{CE}}(\mathbf{e}^r_c),
    \end{align}
    where $\lambda$ controls the contribution of $\mathcal{L}_{\text{CE}}$. 
    
     \textbf{Seg}: To simulate autoregressive training, the entire utterance is split into two segments. The extraction of the second segment incorporates past extracted speaker embedding as input. Following the settings in NeuroHeed \cite{pan2024neuroheed}, we define the window length $L_\text{win} \sim \mathcal{U}(1.05 \text{s}, 3.2 \text{s})$, shift length $L_\text{shift} \sim \mathcal{U}(0.05 \text{s}, 0.2 \text{s})$, and initialization length $L_\text{init} \sim \mathcal{U}(0.05 \text{s}, 3 \text{s})$, where $\mathcal{U}$ denotes a uniform distribution, and the unit is in seconds (s).
In the first window step, MoMuSE processes the mixed audio and visual sequence from the first segment:
\begin{equation}
\hat{\mathbf{x}}(1), \mathbf{e}^r_c(1) = \text{MoMuSE}(\mathbf{y}(1), \mathbf{v}(1)).
 \vspace{-1mm}
\end{equation}
The memory bank is initialized as per Eq.~\ref{eq:eq1}.
In the subsequent window step, MoMuSE utilizes stored voiceprint information from the memory bank to estimate the second segment’s speech and generate momentum-based speaker embedding sequences:
\begin{equation}
\hat{\mathbf{x}}(2), \mathbf{E}^r_m(2) = \text{MoMuSE}(\mathbf{y}(2), \mathbf{v}(2), \mathbf{e}^r_a).
\end{equation}
To enhance the accuracy of the momentum-based speaker embedding, we define the segment loss as:
\begin{equation}
\mathcal{L}_{\text{seg}} = \mathcal{L}_{\text{SI-SNR}}(\mathbf{x}(2), \hat{\mathbf{x}}(2)) + \lambda \sum^R_{r=1} \mathcal{L}_{\text{CE}}(\overline{\mathbf{E}}^r_m(2)),
 \vspace{-1mm}
\end{equation}
where $\overline{\mathbf{E}}^r_m(2) \in \mathbb{R}^{H \times 1}$ is the mean of $\mathbf{E}^r_m(2)$ across the time dimension.

The total loss is designed as follows: 
\begin{equation}
    \mathcal{L}_{\text{total}} = \alpha \mathcal{L}_{\text{utt}} + \beta \mathcal{L}_{\text{seg}} + \gamma \mathcal{L}_{\text{pe}},
     \vspace{-1mm}
\end{equation}
where $\alpha$, $\beta$, and $\gamma$ are hyperparameters balancing the contribution of each loss component



\section{Experimental Details}
 \vspace{-1mm}

\subsection{Dataset}

All the models were trained on the Voxceleb2 dataset \cite{chung18b_interspeech}, and the dataset simulation scripts in \cite{pan2021muse} were adopted to generate our dataset. The simulated 2-speaker mixed speech dataset contains 20k, 5k and 3k utterances for the training, validation and test sets, respectively. The Signal-to-Noise (SNR) ratio was randomly chosen from $-10$ to $10$ dB.  
 The videos were sampled at 25 FPS and the  synchronized audio was sampled at 16 kHz.
For each utterance, a random type of visual impairment from the set $\{$visual missing, lip concealment, low resolution$\}$ was applied. The impairment ratio was randomly chosen from [0\%, 80\%) for the training and validation sets, and  from [0\%, 100\%) for the test set. 
  \vspace{-1mm}

\subsection{Implementation details}
 
The initial learning rate was set to $1e{-3}$  when optimizing from scratch and $1e^{-4}$ when initialized from a pre-trained checkpoint, using the Adam optimizer. 
The learning rate was halved if the best validation loss (BVL) did not improve within six consecutive epochs, and the training stopped if the BVL did not improve within ten consecutive epochs. The maximum number of training epochs was set to 100. 
We set 
$\theta$ to a relatively large value 0.7 to make sure $\mathbf{e}^r_a$ was replaced only when  $\mathbf{e}^r_c$ was accurate enough.  For online inference, we set $L_\text{init}$, $L_\text{win}$ and $L_\text{shift}$ to 1s, 2.7s and 0.2s, respectively, according to  \cite{pan2024neuroheed}. The hyperparameters  $\alpha$,  $\beta$,  $\gamma$, and $\lambda$  were set to 0.3, 0.7, 0.05, and 0.1, respectively.

The SI-SNR loss is scale-invariant, meaning the energy of the estimated speech varies across different window steps. To ensure consistent energy levels in the speech, we applied the NeuroHeed normalization strategy \cite{pan2024neuroheed} during inference.
\section{Results}
\subsection{Causal vs. Non-causal}
In this section, we explore the optimal training settings for online scenarios where visual cues are impaired.
\renewcommand{\arraystretch}{0.9}
\begin{table}[htbp]
	\centering
	\caption{ SI-SNR (dB) of MuSE with different training settings}
        \vspace{-2mm}
	\label{tab:causal}
	\begin{tabular}{c c|c|c|c|c}
		\midrule
            \multicolumn{2}{c|}{Training setting} & \multicolumn{4}{c}{Evaluation mode}\\  
		 \multirow{2}{*}{Causal}& \multirow{2}{*}{Impaired} & \multicolumn{2}{c}{Offline}& \multicolumn{2}{c}{\textbf{Online}}\\ \cmidrule{3-6}
        
         & & Normal & Impaired &  Normal & \textbf{Impaired}   \\ \midrule  
         \XSolidBrush & \XSolidBrush \cite{pan2021muse} &  \textbf{11.70}  & 7.08 & \textbf{9.08} & 3.67 \\
         \XSolidBrush & \Checkmark &  10.11 & \textbf{11.30} & 8.74  &\textbf{6.05} \\  
         \Checkmark & \XSolidBrush & 9.73 &5.73 &7.49 &2.69 \\ 
         \Checkmark & \Checkmark &9.34 &8.26 &7.20 & 4.69 \\ 
		\midrule
	\end{tabular} 
    \vspace{-2mm}
\end{table}

 Table \ref{tab:causal} demonstrates that the non-causal model configuration \cite{luo2019conv} outperforms the causal variant in both online and offline evaluation modes.
Additionally, incorporating impaired video data into the training set significantly enhances performance in scenarios where visual cues are unreliable. However, this improvement comes at the cost of reduced performance when the visual inputs are normal. This trade-off likely occurs because the model learns to rely less on visual information when it is exposed to impaired visual data during training.
The second row of Table \ref{tab:causal}, which employs a non-causal model and incorporates impaired data during training, shows the best performance in our target scenario, online inference with impaired visual cues. Therefore, this configuration is adopted in subsequent experiments to ensure optimal results.

\vspace{-1mm}
\subsection{Comparative Analysis with Baselines}
We evaluate our proposed MoMuSE model and several baseline approaches using multiple performance metrics: Scale-Invariant Signal-to-Noise Ratio (SI-SNR), Signal-to-Distortion Ratio (SDR) \cite{vincent2006performance}, Perceptual Evaluation of Speech Quality (PESQ) \cite{rix2001perceptual}, and Short-Time Objective Intelligibility (STOI) \cite{taal2011algorithm}. The results are summarized in Table \ref{tab:baseline}.
\begin{table}[htbp]
    \centering
    \vspace{-5mm}
    \caption{`V' and `I' denote visual sequences and static image, respectively. `Data' denotes the number of utterances in  training set  }
    \vspace{-2mm}
    \begin{tabular}{c|c|c|c|c|c|c}
    \midrule
    Model & Cue & SI-SDR & SDR & PESQ &STOI &Data\\ \midrule
         FaceFilter \cite{chung20c_interspeech}&I&-& 2.5 &- & -&2,000k  \\ 
         VisualVoice \cite{gao2021visualvoice} &I &- & 7.06 & -&\textbf{0.80}& 1,000k   \\
        MuSE & V&  6.05 & 6.66 & 1.67 & 0.77&20k \\ 
         MoMuSE &V &\textbf{6.87} & \textbf{7.59} & \textbf{1.81}&0.78 &20k\\ \midrule
    \end{tabular}
    \vspace{-2mm}
    \label{tab:baseline}
\end{table}

First, compared to baselines that utilize a static image (I) as the visual cue (e.g., FaceFilter and VisualVoice), MuSE and MoMuSE leverage dynamic visual sequences (V), achieving comparable or even superior performance while requiring less training data. This highlights the effectiveness of using time-varying visual features (such as lip movements) to enhance speech separation. 
Second, when visual cues are unreliable or absent, the use of  historical voiceprint stored in memory bank as complementary information boosts performance. MoMuSE outperforms MuSE across all evaluation metrics, showing its robustness in real-world scenarios with impaired visual cues.

\subsection{Ablation Studies}

\begin{figure}[htbp]
    \centering
    \vspace{-3mm}
    \includegraphics[width=0.38\textwidth]{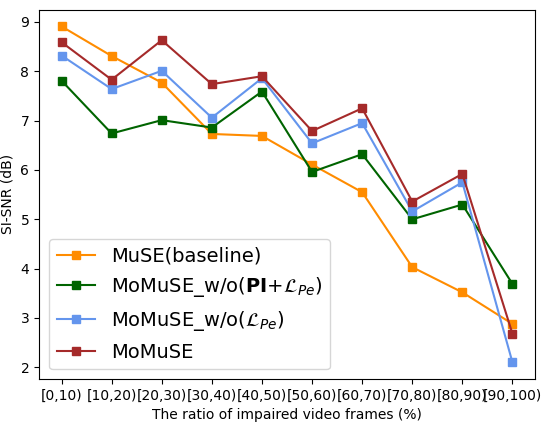}
    \vspace{-3mm}
    \caption{Model performance under varied impaired ratios is shown, where MoMuSE\_w/o(\textbf{PI}+$\mathcal{L}_{\text{Pe}}$) excludes  parameter initialization (\text{PI}) and penalty loss ($\mathcal{L}_{\text{Pe}}$), and  MoMuSE\_w/o($\mathcal{L}_{\text{Pe}}$) excludes only $\mathcal{L}_{\text{Pe}}$ }
    \label{fig:ratio__}
    \vspace{-2mm}
\end{figure}
Fig.\ref{fig:ratio__} illustrates the impact of different training strategies on model performance across varying levels of visual impairment. Compared to MoMuSE\_w/o($\mathcal{L}_{\text{Pe}}$), MoMuSE demonstrates improved performance across all scenarios, highlighting the importance of the $\mathcal{L}_{\text{Pe}}$ loss. By prioritizing the current speaker embedding $\mathbf{e}^r_c$, this loss function ensures more discriminative and effective representations, particularly in scenarios with inconsistent or absent visual cues.

\begin{table}[htbp]
\vspace{-2mm}
	\centering
	\caption{Comparative analysis of various types of impairments}
        \vspace{-2mm}
	\label{tab:impairments}
	\begin{tabular}{c|c|c|c|c|c}
		\midrule
		 Impairments &Model   & SI-SNR   & SDR  & PESQ & STOI \\ \midrule
        
         Visual & MuSE &3.46 &4.19 &1.53 & 0.71\\ 
         Missing& MoMuSE &\textbf{6.13} &\textbf{6.92} &\textbf{1.79} &\textbf{0.76} \\ \midrule

         Lip 
        &MuSE & \textbf{ 7.32}&\textbf{7.88} &1.73 &\textbf{0.80} \\ 
          Concealment&MoMuSE&7.08 &7.78 &\textbf{1.81} & 0.78\\ \midrule

          Low 
          & MuSE & 7.37&7.92 &1.75 &\textbf{0.80} \\ 
        Resolution & MoMuSE &\textbf{7.39} &\textbf{8.06} &\textbf{1.84} & 0.79\\ \midrule
			\end{tabular} 
   \vspace{-2mm}
\end{table}
Additionally, the Parameter Initialization (PI) strategy further enhances performance by initializing the model with pre-trained parameters, which strengthens the quality of learned embeddings and improves adaptability to varying impairment levels. MoMuSE also outperforms MuSE, especially in high visual impairment scenarios (over 20\%), by effectively leveraging historical voiceprint information as a fallback when visual cues are unreliable. However, a slight performance drop is observed in low impairment scenarios, likely due to reduced reliance on visual cues.

Although MoMuSE exhibits strong overall performance, its effectiveness decreases as the impairment ratio increases, suggesting that visual cues remain more reliable than voiceprint-based information when both modalities are available.

\vspace{-1mm}
\subsection{Impact of the Impairment Types}

Table \ref{tab:impairments} presents the performance of MoMuSE and MuSE on three types of visual impairments. 
In scenarios of complete visual absence, where no useful visual information is available, MoMuSE shows clear effectiveness, highlighting its ability to excel when visual cues are severely compromised. This result underscores the model's strength in leveraging momentum-based speaker embeddings to maintain robust performance. For cases of lip concealment and low resolution, although MoMuSE does not exhibit significant improvements, its performance remains competitive, suggesting that even with partial visual impairments, the model can effectively utilize available cues while relying on speaker momentum for stability.

\vspace{-1mm}
\subsection{Analyzing Total Visual Absence}

In this section, we evaluate the model’s performance in a challenging scenario where visual cues are available only during the initialization step (the $1^{st}$ second) and are entirely absent in subsequent steps.
\begin{table}[htbp]
\vspace{-3mm}
    \centering
    \caption{Comparative analysis on total visual absence. The performance of models on different utterance lengths.}
    \vspace{-2mm}
    \label{tab:length}
    \begin{tabular}{c|c|c|c|c|c}
        \midrule
        \diagbox{}{} & [0s, 5s) & [5s, 10s)  & [10s, 15s) & [15s, 20s) & [20s, $\infty$) \\
        \midrule
       MuSE & \cellcolor{gray!54}3.44 & \cellcolor{gray!42}1.23 & \cellcolor{gray!28}-1.38 & \cellcolor{gray!31}-0.90 & \cellcolor{gray!0}-6.81 \\
        MoMuSE & \cellcolor{gray!74}7.24 & \cellcolor{gray!77}7.81 & \cellcolor{gray!76}7.69 & \cellcolor{gray!90}10.88 & \cellcolor{gray!100}12.25 \\
        \midrule
    \end{tabular}
    \vspace{-2mm}
\end{table}

Table \ref{tab:length} presents the SI-SNR (dB) performance of models with varying utterance lengths. First, MoMuSE consistently outperforms MuSE, demonstrating its ability to maintain focus on the target speaker over time in the absence of visual cues. Second, as the length of the utterance increases, MoMuSE shows improved performance. This improvement can be attributed to its updating mechanism, which discards poorer speaker embeddings and retains more reliable ones in the memory bank during inference. Using these better speaker embeddings, MoMuSE is able to achieve enhanced performance.

An interesting observation is  MoMuSE’s performance in Table \ref{tab:length} exceeds its performance in Table \ref{tab:impairments}. We attribute this to the clean visual input during the initialization step in Table \ref{tab:length}, whereas impaired visuals might occur in the initialization step in Table \ref{tab:impairments}. This highlights the critical importance of having clean visuals during the initialization step.

\vspace{-2mm}
\section{Conclusion and Future Work}
\vspace{-1mm}


We have proposed the MoMuSE, an extension of the MuSE model with a momentum mechanism, to track the target speaker in AV-TSE tasks even when visual cues are impaired or missing. Our approach uses a built-in memory bank to maintain the target speaker's hidden state, eliminating the need for audio pre-enrollment. In the future, we will focus on refining the confidence mechanism and memory bank design to adapt to diverse scenarios.
\vspace{-2mm}

\bibliographystyle{IEEEbib}
\bibliography{ref}

\end{document}